# Strong electromagnetic pulses generated in high-intensity short-pulse laser interactions with thin foil targets

P. Rączka[1*], J.-L. Dubois[2], S. Hulin[2], V. Tikhonchuk[2]
M. Rosiński[1], A. Zaraś-Szydłowska[1], and J. Badziak[1]

[1] *Institute of Plasma Physics and Laser Microfusion, Warsaw, Poland*

[2] *CELIA, University of Bordeaux-CNRS-CEA, Talence, France*

**Abstract**

Measurements are reported of the target neutralization current, the target charge, and the tangential component of the magnetic field generated as a result of laser-target interaction by pulses with the energy in the range of 45 mJ to 92 mJ on target and the pulse duration from 39 fs to 1000 fs. The experiment was performed at the Eclipse facility in CELIA, Bordeaux. The aim of the experiment was to extend investigations performed for the thick (mm scale) targets to the case of thin (μm thickness) targets in a way that would allow for a straightforward comparison of the results. We found that thin foil targets tend to generate 20%-50% higher neutralization current and the target charge than the thick targets. The measurement of the tangential component of the magnetic field had shown that the initial spike is dominated by the 1 ns pulse consistent with the 1 ns pulse of the neutralization current, but there are some differences between targets of different type on sub-ns scale, which is an effect going beyond a simple picture of the target acting as an antenna. The sub-ns structure appears to be reproducible to surprising degree. We found that there is in general a linear correlation between the maximum value of the magnetic field and the maximum neutralization current, which supports the target-antenna picture, except for pulses 100's of fs long.

*/ Corresponding author: P. Rączka, Division of Laser Plasma, Institute of Plasma Physics and Laser Microfusion, ul. Hery 23, 01-497 Warsaw, Poland; tel. +48 22 6381005 ext. 20.

1.INTRODUCTION

One of the characteristic effects observed in the experiments performed with the use of high-energy, high-intensity lasers is the appearance of strong electromagnetic pulses (EMP) with frequencies ranging from tens of MHz to multi-GHz. First accounts of the rf to microwave emission resulting from the laser-target interactions were published already in the seventies (Pearlman & Dahlbacka, 1978). With the advent of petawatt lasers and MJ laser facilities the issue of EMP became of considerable practical interest, because such pulses strongly interfere with the electronics used to collect data and manipulate targets and hence pose a serious threat to safe and reliable execution of experiments. Therefore a dedicated effort was made to study the EMP effect at facilities such as Vulcan, Titan, Omega, NIF, LMJ and other (Mead et al., 2004; Raimbourg, 2004; Stoeckl et al., 2006; Remo et al., 2007; Brown et al., 2008; Bourgade et al., 2008; Eder et al., 2009; Brown et al., 2010; Eder et al., 2010; Chen et al., 2011; Bateman & Mead, 2012; Brown et al., 2012; Brown et al., 2013;). Pulses of 100's ns duration and electric fields of 100's kV/m strength were recorded. Various EMP generation mechanisms had been considered at this stage, including the charge separation effects in laser plasmas and low-frequency oscillations of the expanding plasmas (Pearlman & Dahlbacka, 1978), the electron currents within the laser plasmas (Felber, 2005), excitation of the chamber eigenmodes by a stream of fast electrons ejected from the target (Mead et al., 2004; Eder et al., 2009), and EMP generation through direct impact of charged particles and radiation (Stoeckl et al., 2006). However, a complete quantitative understanding of this phenomenon is still lacking. The EMP effect continues to attract considerable attention and is now being investigated also with the use of smaller laser facilities (Aspiotis et al., 2006; Miragliotta et al., 2011; Consoli et al., 2013; Dubois et al., 2014; De Marco et al., 2014; Cikhardt et al., 2014; Varma et al., 2014; Poyé et al., 2015a; Poyé et al., 2015b; Price et al., 2015; Consoli et al., 2015a, 2015b, 2016; De Marco et al., 2016, 2017; Yi et al., 2016; Robinson et al., 2017; Krása et al., 2017a). In particular, a series of recent articles (Dubois et al., 2014; Poyé et al., 2015a; Poyé et al., 2015b) concentrated on the effect of electromagnetic emission related to the target neutralization current, which is of pulsed nature and which appears as a consequence of the laser-induced electric polarization of the target (Pearlman & Dahlbacka, 1977; Benjamin et al., 1979; Beg et al., 2004; Quinn et al., 2009; Krása et al., 2015; Krása et al., 2017b). An extensive experimental and theoretical effort was undertaken to study this effect. Several experiments were performed at the

Eclipse laser in CELIA, Bordeaux, in which thick (~3 mm) targets of various size and made of various metals and dielectrics were irradiated with laser pulses of the energy in the range 30-100 mJ on target and the duration 30-1000 fs. A special form of the target called a "lollipop" was used, which facilitated the measurement of the neutralization current. The "lollipop" target was mounted on a thin grounded metal plate which ensured the presence of a mirror charge and hence the dipole character of the radiation pattern. The target charge and the maximum value of the current were found to scale in direct proportionality to the pulse energy, and to be only weakly dependent on the pulse duration (Dubois et al., 2014). The spectrum of the emitted electromagnetic signal recorded by a B-dot magnetic probe had a peak at 1 GHz, consistently with the 1 ns time scale of the return current (Poyé et al., 2015a). For moderate values of the target charge (up to 10 nC) the amplitude of the magnetic field oscillations was found to scale in direct proportionality to the charge, which is consistent with a simple formula for the dipole radiation in the far field.

It is of obvious interest to extend the analysis of (Dubois et al., 2014; Poyé et al., 2015a) to the case of very thin (micrometer scale) foil targets, such as those commonly used in the laser-driven ion acceleration experiments using short pulse laser. To this end we performed an experiment at the Eclipse laser using custom made targets which had the "lollipop" form, but were at the same time capable of supporting very thin foils, thus giving us the opportunity to make a meaningful comparison of the charging and discharging of thick and thin targets. Several electromagnetic probes and particle diagnostics were used in this experiment, but in this note we report only the results on the neutralization current, the target charge and the tangential component of the magnetic field, deferring other results to a subsequent article.

2. EXPERIMENTAL SETUP

Our experiment was performed at the Eclipse facility in CELIA, Bordeaux. Eclipse is a Ti:Sa laser delivering a linearly polarized beam with 807 nm central wavelength, focused by an f/5 off-axis parabolic mirror to the focal spot with the FWHM diameter of 10.50 μm. The laser beam was incident on the target along the target normal. The pulse duration was varied in the range 39 fs – 1000 fs, with the uncertainty of 2 fs for the shortest pulse, increasing to 15 fs for the 200 fs pulse and then leveling off to 20 fs for 1000 fs pulse. The laser pulse energy was varied in the range 99 mJ – 212 mJ with 2% rms jitter, which gave 45 mJ – 95 mJ on target with 5% rms overall uncertainty after taking into account (45±2)% compression efficiency. The ns pedestal due to the amplified spontaneous emission (ASE) was found to be of 5 ns duration. The contrast for the shots 60-184 – which constitute the majority of data analyzed in this article – was $2\times10^{-6}$ at 150 ps before the peak, and for the shots 41-59 it was of $6\times10^{-6}$ @150 ps. We used the Sophia experimental chamber, which is a cylinder 520 mm high, with 1000 mm internal diameter, topped by a flat concave cap.

Targets used in our experiment were placed in a ''lollipop'' target holder which consisted of a brass ring 14.0 mm in diameter, supported on a thin brass wire stalk 28.0 mm long. Inside the brass ring the following targets were placed: (a) a massive Cu "pill" 10.1 mm in diameter and 1.0 mm thick; (b) an Al foil 6.0 μm thick, with 0.3 μm layer of polystyrene on the rear side, further denoted as "AlPs", pasted on the rear side of a Cu pill 10.1 mm in diameter and 1.0 mm thick, with 10 holes 1 mm in diameter for shots, as shown in Fig. 1 (the expectation was that such the polystyrene layer would increase the number of the TNSA protons, thus making them easier to detect and characterize); (c) pure Al foil 6.0 μm thick, placed between two Cu pills, each 10.1 mm in diameter and 0.50 mm thick, with 10 holes 1 mm in diameter for shots. It was expected that targets of the type (a) would give results close to those reported in (Dubois et al., 2014; Poyé et al., 2015a), despite slightly smaller thickness (1 mm compared to 3 mm); the targets of the type (b) and (c) should then clearly demonstrate the differences in the target charge and EMP generation from thick targets and thin foil targets.

The target holder was mounted on a thin square Al plate (160.5 mm×160.5 mm and 4.2 mm thick) that was electrically grounded. The target holder was directly connected via a 50 ohm SMA mount to a coaxial cable, which allowed for a direct measurement of the target neutralization current. Several probes were placed inside and even outside the chamber to monitor the electromagnetic field and obtain a good characterization of the generated EMP. Signals registered by the electromagnetic probes were recorded using two high-bandwidth oscilloscopes, which were externally triggered by a photodiode. There was no electromagnetic shielding for the oscilloscopes, but we verified that the signal pickup over air with cables disconnected from the chamber was negligible. In this article we report only the results for the target neutralization current, the target charge, and the measurements of the tangential component of the magnetic field performed with two B-dot magnetic field probes. A schematic view of the experimental chamber with the indicated location of the B-dot probes is shown in Fig. 2.

3.NEUTRALIZATION CURRENT AND THE TOTAL TARGET CHARGE

In order to measure the target neutralization current the target stalk was directly connected via a coaxial cable to a high-bandwidth oscilloscope Teledyne LeCroy SDA 760Zi-A (6 GHz bandwidth, 20 GSa/s) In most of the shots a 66 dB external attenuation on the cable was used. The results on the neutralization current and the total target charge reported below come from the shots 42-59, 104-125, 127-145, 166-178 (shots 1-41 were devoted to adjustments in the experimental setup and data collection procedures). The temporal dependence of the target neutralization current for three shots that are representative of the whole data sample is shown in Fig. 3. Apart for the value of the maximal current the curves are very similar for all three types of targets considered here, with the dominant initial pulse having a characteristic "knife blade" shape, followed by several oscillations around the zero point, which are strongly damped. The time scale of the initial pulsation is ~1 ns.

In order to extract the value of the target charge the following procedure was used: (1) the data preceding the maximum by more than 5 ns (i.e. the duration of the ASE pedestal) was averaged to identify the offset of the zero point for the measurement of the current; (2) the data was corrected for the zero point offset and all the data for times preceding the maximum current by more than 5 ns was excluded from the analysis; (3) the properly subtracted data for the neutralization current was numerically integrated using the trapezoidal rule, giving the target charge as a function of time, as shown in Fig. 4; (4) the estimate for the target charge from such plots was obtained by taking an average of the charge at the second maximum and second minimum. The plot of the target charge as a function of the laser energy on target, for the pulse duration of approximately 40 fs, is shown in Fig. 5. For the laser parameters for which three or more consistent data points were obtained only the averages are presented, together with the error bars, and the rest of the data is shown in the form of a scatter plot. We find that the measurements for the thick Cu target (shots 127-145) are consistent with the direct proportionality of the total target charge to the laser pulse energy, as it was observed in (Dubois et al. 2014; Poyé et al. 2015a). Surprisingly, however, the numerical values of the charge that we obtain for the thick target in this experiment are visibly higher: from the Fig. 6 in (Dubois et al. 2014) we find that the target charge for the 50 fs pulse duration and the pulse energy 88 mJ is approximately (12±1) nC, whereas in our measurement we find for the 39 fs pulse duration and the pulse energy 85 mJ the value (27±2) nC. This surprising difference cannot be attributed to the shorter duration of the pulse, as may be seen from the Fig. 8a in (Poyé et al. 2015a), where data for the 30 fs pulse duration are displayed. Our view is that this difference may be attributed to the difference in the laser contrast: the measurements reported in (Dubois et al. 2014; Poyé et al. 2015a) were performed with the laser contrast of $1\times10^{-7}$ and 2 ns duration of the ASE pedestal, whereas for the shots 127-145 on the thick Cu target the contrast was $2\times10^{-6}$@150 ps and the duration of the ASE pedestal was 5 ns. This view is supported by the measurement of the target charge for the shots 6-12 on the thick Cu target, for which the contrast was $7\times10^{-5}$@140 ps and for which the value of $(35\pm2)$ nC is was obtained. This indicates that the degree of electric polarization of the target is quite sensitive to the amount of preplasma, which is an observation that certainly deserves a more detailed investigation in further experiments.

Turning to thin targets, we see in Fig. 5 that the charge deposited on the pure Al thin foil target (shots 42-56) appears to be approximately 50% bigger than the charge on the thick Cu target. The charge deposited on the thin AlPs target is also larger than the charge on the thick Cu target, but by a smaller factor, approximately 20%. The fact that very thin targets generate higher charge is natural, since the outflow of the electron from the target occurs not only on the front side, but also on the rear side, where fast electrons penetrate through the target and form a sheath. The fact that there is a difference between Al and AlPs targets may be due to two factors: firstly, the layer of Ps on the rear of the target - introduced with the aim to enhance the number of the laser accelerated protons - may reduce at the same time the number of escaping electrons; secondly, the ns contrast for the shots with pure Al foil was $8\times10^{-6}$ at 150 ps instead of $2\times10^{-6}$ for the shots on the AlPs foil (which again indicates the possible importance of the preplasma).

In Fig. 6 we show the target charge as a function of the pulse duration, for the pulse energy close to 90 mJ. Also in this case we find the charge generated on a thin foil target to be noticeably higher than the charge on a thick Cu target by approximately 30% for longer durations of the pulse. For both types of target the charge falls by approximately 20% when the pulse duration is varied from 40 fs to 500 fs. However, some data points obtained for the AlPs foil at 60 fs suggest that the charge generated for this pulse duration is in fact larger than that obtained for the shortest pulse, which is counterintuitive. It would be interesting to explore this effect in more detail with better statistics and to check whether this might correlate with somewhat higher energies of the ions accelerated from this foil, which would be important for ion acceleration using high power fs lasers.

Given the similarity in the temporal shape of the neutralization current the most relevant parameter for the target polarization is thus the maximum value of the current. In Fig. 7 we show the maximum value of the neutralization current as a function of the pulse energy, for the pulse duration close to 40 fs. We find similar trends as in the case of the target charge. The values obtained for the thick Cu target are consistent with the direct proportionality between the maximum current and the pulse energy, as observed in (Dubois et al. 2014), and the values measured for the Al and AlPs targets are higher than for the thick Cu target. It should be noted that the maximum value of the neutralization current obtained for the thick Cu target in this experiment is somewhat bigger than that reported in (Dubois et al. 2014; Poyé et al. 2015a) for similar laser parameters as in the previous experiment at Eclipse laser (although incidentally it appears to be consistent with the simulation results displayed in Fig. 13.b. of Dubois et al. 2014). The maximum value of the return current for thin Al foil targets is in turn approximately 50% bigger than the maximum value for the thick Cu target. In Fig. 8 we show the maximum current as a function of the pulse duration, for the energy close to 90 mJ. Again, the trends are similar to those observed for the target charge.

## 4. MEASUREMENTS OF THE MAGNETIC FIELD

In the following we report measurements of the tangential component (referring to natural cylindrical coordinates in the cylindrical Sophia chamber) of the magnetic field using a Prodyn RB-230 probe, with an average equivalent area of approximately $3\times10^{-5}$ m$^2$ and the practical operation range limited to frequencies above 400 MHz, further denoted as B-dot1, and a Prodyn B24(R) probe with an average equivalent area of approximately $9\times10^{-6}$ m$^2$, further denoted as B-dot2, used in this measurement mainly for cross-check. Each of those probes was connected to a Prodyn BIB-100G balun to provide an unbalanced, symmetrized signal. Each balun introduced 8 dB attenuation of the signal. The B-dot1 probe was located 214 mm behind the target, 41 mm above the target, and 55 mm to the right of the target, and was connected by a 2 m coaxial cable to the Teledyne LeCroy oscilloscope. In order to determine the physical value of the $B_{tg}$ the signal from the B-dot probes was processed in the following way: (a) the raw signal was corrected for the external attenuation (0 or 6 dB) and for the attenuation of the balun (8 dB, assumed to be independent of the frequency); (b) a band pass fast Fourier Transform (FFT) filter was used to eliminate frequencies below 400 MHz (which are not properly recorded by the probe B-dot1), as well as spurious high frequency contribution (with the 20GSa/s sampling rate of the oscilloscope the spectrum of the discrete Fourier Transform extends to 10 GHz); (c) the correction for the frequency-dependent attenuation of the coaxial cable was applied; (d) the inverse FFT was performed to obtain corrected and regularized data for $dB_{tg}/dt$; (e) the $B_{tg}$ was obtained as a function of time by direct integration of $dB_{tg}/dt$. The $B_{tg}$ signal has the form of an initial strong spike followed by damped oscillation which persist for about 200 ns, as is shown in Fig. 9 for three shots representative of a broader sample: a thin Al foil (shot 49), a thin AlPs foil (shot 110) and a thick Cu target (shot 145). The displayed shots correspond to similar values of the laser parameters, i.e. the pulse energy close to 90 mJ and the pulse duration approximately 40 fs. The EMP generated from the thin AlPs target is stronger than the EMP from the thick Cu target, while the EMP from the thin Al target is stronger than EMP from the thin AlPs target – which is not very surprising if we take into account our previous discussion of the neutralization current and the target charge. It is these initial spikes which are most interesting, because there is some universality in their appearance, since they are a direct signature of the laser-target interaction and cannot be affected by other factors that may result in the generation of the EMP in the experimental chamber. The short time-scale structure of the initial spikes for those three shots is shown in Fig. 10. In a first approximation these are oscillations with a period of ~1 ns, which is consistent with the characteristics of the target neutralization current. A closer analysis reveals however the presence of modulations of a much shorter time scale. This is best seen by looking at $dB/dt$ instead of $B$ itself, as shown in Fig. 11. This picture may look random and chaotic, but interestingly it is reproduced from shot to shot with surprising accuracy. To underscore this we show in Fig. 12 the same plot for shots at two different targets with the pulses of 500 fs duration. The similarity between Fig. 12 and Fig. 11 is stunning, given the fact that they correspond to laser pulses differing in duration by an order of magnitude. The fine structure of the initial spikes corresponds to frequencies visibly higher than 1 GHz and hence it must be produced by EMP generation mechanisms different from the simple target neutralization. We also note that there are qualitative differences in the signal patterns for targets of different structure (thick Cu, thin Al and thin AlPs), which is consistent with the conclusion that part of the signal is not directly related to the target neutralization.

The results on the measurement of $B_{tg}$ obtained from the B-dot2 were consistent with the results from the B-dot1. The B-dot2 probe was placed at the same position along the laser beam direction as the target, 10 mm below the target and 335 mm to the right of the target. It was connected by two 0.62 m coaxial cables to the BIB-100G balun, and then via a 5.00 m cable to the LeCroy oscilloscope. For the shots 104-113 on a thin AlPs target the B-dot2 probe was set to measure the tangential component of the magnetic field, thus providing a good cross-check on the B-dot1 measurement. The signal from B-dot2 probe was shifted in time so that the initial spikes would overlap in time. We see in Fig.13 that the initial spikes from both probes match very closely, as it should be, given the fact that the distances from the target for both probes are similar, and there is qualitative agreement for later times. It should be noted that the signal of the B-dot2 probe may be affected by the fact that it is placed almost at the level of the ground metal plate supporting the target, so that the mirror charge mechanism ensuring the dipole pattern of radiation from the target may not operate as well as for the B-dot2 probe. Furthermore, at later times the probe records reflections of the initial spike, which for the B-dot2 probe might be different than for the B-dot1.

One may also wonder whether our measurements could be affected by charged particles or X-rays interacting directly with the probes. However, in the case of the initial spike any effect of charged particles is excluded, as they are much slower than light. As for the direct influence of hard radiation, this was thoroughly investigated for the Titan laser (Eder et al., 2009) and was found to have a minor effect.

We conclude our discussion the time-dependence of $B_{tg}$ by showing in Fig. 14 the spectrum of $B_{tg}(t)$ for the shots at three targets discussed above. There are two characteristic features of this spectrum: firstly we see a broad maximum around 1 GHz, which is consistent with the similar feature in the FT of the neutralization current. It should be noted, however, that in the case of thin targets this broad maximum is accompanied by at least three high and narrow spikes: at 0.98 GHz, 1.16 GHz and 1.42 GHz. It is tempting to associate these spikes with excitation of the chamber eigenmodes, so we consider this question here in some detail. The Sophia experimental chamber is approximately a cylindrical cavity; the formula for the eigenfrequencies in such a cavity may be found for example in (Jackson, 1999). The modes are labeled by the azimuthal number $n$ characterizing the number of nodes in the dependence of the fields on the azimuthal angle, the number $m$ characterizing the number of nodes in the radial direction and the number $p$ characterizing the number of nodes in the axial direction. The boundary conditions restrict the modes to have either $B_z = 0$, denoted as TM modes, or $E_z = 0$, denoted as TE modes. Taking into account that the Sophia chamber has the radius 0.50 m and assuming and average height of 0.51 m we find that the four lowest TE modes are $TE_{111}$(343 MHz), $TE_{211}$(414 MHz), $TE_{011}$(469 MHz), and $TE_{311}$(587 MHz), whereas the four lowest TM modes are $TM_{010}$(230 MHz), $TM_{110}$(366 MHz), $TM_{011}$(373 MHz), and $TM_{111}$(469 MHz). For higher values of *m, n, p* there is proliferation of modes and they are rather densely spaced (see the discussion of the eigenmodes in the cylindrical chamber of the Titan laser given in Bateman & Mead, (2012)). Due to the dense distribution of the eigenmodes in the frequency space we are not able at this point to confirm or exclude the association of those peaks with particular modes; this would require a more detailed analysis.

It is of obvious interest to ask what is the maximum value of the EMP magnetic field for the given laser parameters and target type. From the discussion reported above we know that the largest values of the $B_{tg}$ are achieved in the initial spike. We therefore extracted maximum values of $B_{tg}$ for shots of approximate duration 40 fs, performed at various energies (Fig. 15) The pattern we find very closely resembles the pattern for the target charge as a function of energy at 40 fs, shown in Fig. 5. This may be visualized through a correlation plot between the maximum values of $B_{tg}$ and the maximum value of the neutralization current, as shown in Fig. 16; a strong linear correlation between these quantities is evident. This is consistent with the expectation that the spikes in the B-field arise from the target and the stalk acting as a dipole antenna. However, as shown in Fig. 17, for the dependence of the maximum values of $B_{tg}$ on the pulse duration (at the energy close to 90 mJ) the situation is different: we find that for thin foil targets the longer pulses tend to generate disproportionately strong initial spikes. This is underscored by the plot of a correlation between maximum value of $B_{tg}$ and the maximum value of the current for this set of shots, as shown in Fig. 18. We clearly see a deviation from a direct proportionality, which may be indicative of a process that goes beyond the dipole antenna mechanism. We note in passing that the maximum values of $B_{tg}$ obtained in our experiment in the case of thick targets are somewhat smaller than those reported in (Poyé et al. 2015a). However, we were informed (Dubois, 2017) that the results of (Poyé et al. 2015a) are being reanalyzed.

In order to estimate what values of the electric field may be involved in these processes we may use the plane wave $E \approx cB$, which should provide a good approximation for the initial spike. A typical value of the magnetic field in our measurements is $3 \times 10^{-5}$ T, for which we obtain from the plane wave relation the electric field strength of 9 kV/m at the distance of 0.23 m from the target.

5. CONCLUSIONS

Summarizing, we performed measurements of the target neutralization current and we determined the target charge for the laser energies in the range of 45 mJ to 92 mJ and laser pulse durations from 39 fs to 1000 fs for a 1 mm thick Cu target, 6 μm thick Al target with 0.3 μm layer of polystyrene, and a pure 6 μm thick Al target. Our measurements allowed for a straightforward comparison of the results for the thick target and the thin foil targets. In the case of the thick Cu target we found the values of the target charge and the maximum neutralization current to be visibly higher than those reported in (Dubois et al. 2014; Poyé et al. 2015a); as a possible explanation of this difference we see the effect of the laser contrast, i.e. that for a worse laser contrast, that leads to the generation of bigger preplasma, the target charge is higher. This dependence is an interesting point to be investigated in future experiments. Comparing the thin 6 μm foil targets with the thick Cu targets we observe that they tend to generate 20%-50% higher neutralization current and the target charge.  A model to explain this effect in a quantitative way is yet to be devised. We also measured the tangential component of the magnetic field for the EMP generated at various laser parameters and various targets. We found that the magnitude of the magnetic field in the initial spike is in general well correlated with the value of the maximum neutralization current, which is consistent with the picture of the EMP being generated by the target acting as a dipole antenna, as was discussed in (Dubois et al. 2014; Poyé et al. 2015a), with longer pulses being a surprising exception We noticed that the initial spikes have a characteristic sub-ns structure which depends on the target type and which appears to be reproducible to astonishing degree. These effects go beyond the dipole antenna emission mechanism.  Explaining this structures may be useful in advancing our understanding of the laser-target interactions.

ACKNOWLEDGMENTS

This research is supported by the Polish National Science Centre grant Harmonia 2014/14/M/ST7/00024. Access to the Eclipse facility was made possible by the support obtained from LASERLAB-EUROPE, European Union's Horizon 2020 research and innovation programme, under grant agreement no. 654148,  project CNRS-CELIA002294. Our collaboration

<p>
<p>has profited greatly from the intellectual environment created by the COST Action MP1208. We are grateful to D. Neely (RAL) for lending us the Prodyn B-24 B-dot probe.

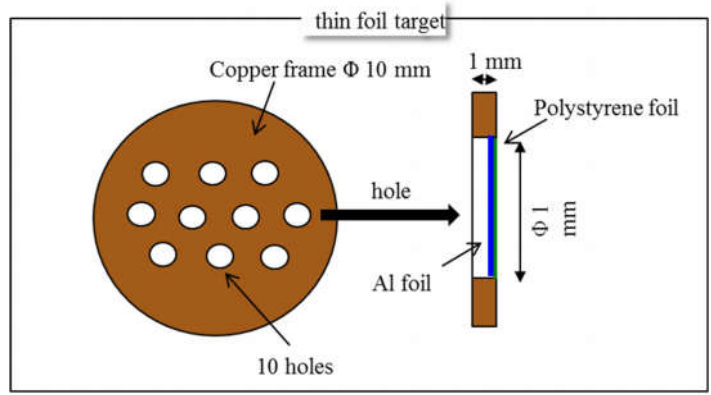

Fig. 1 A 1 mm thick and 10 mm in diameter Cu "pill" with 1 mm holes in which very thin foils could be pasted, allowing for a straightforward comparison of the EMP generated from a 1 mm thick Cu target and a 6 μm thin Al foil.

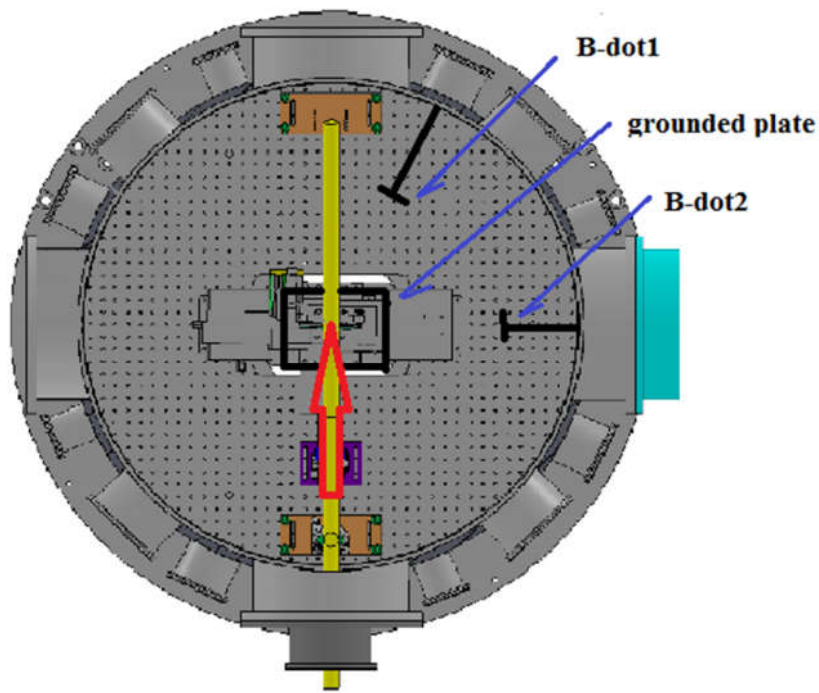

Fig. 2 A schematic view of the experimental chamber, with the indicated location of the B-dot probes.

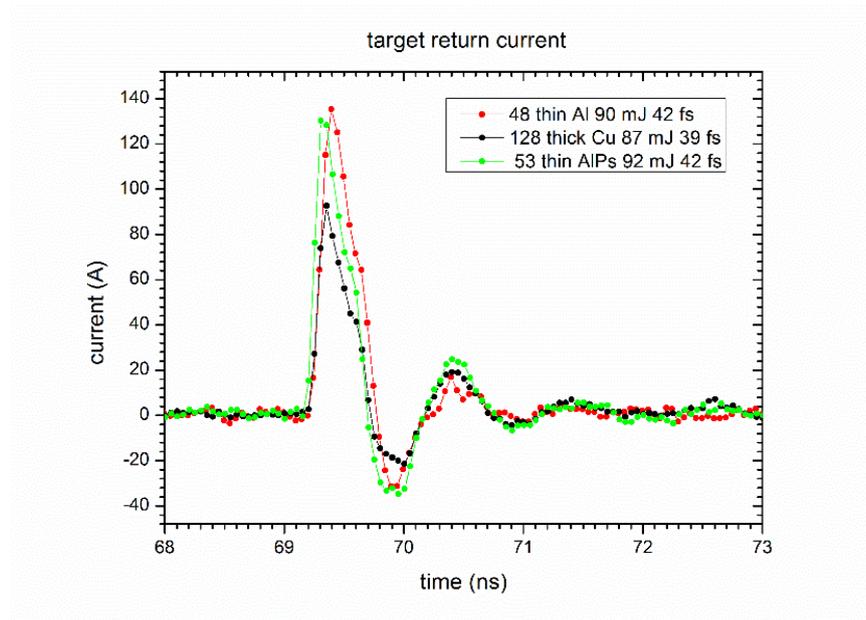

Fig. 3 The target neutralization current as a function of time for three shots at different types of targets, representative of a bigger sample.

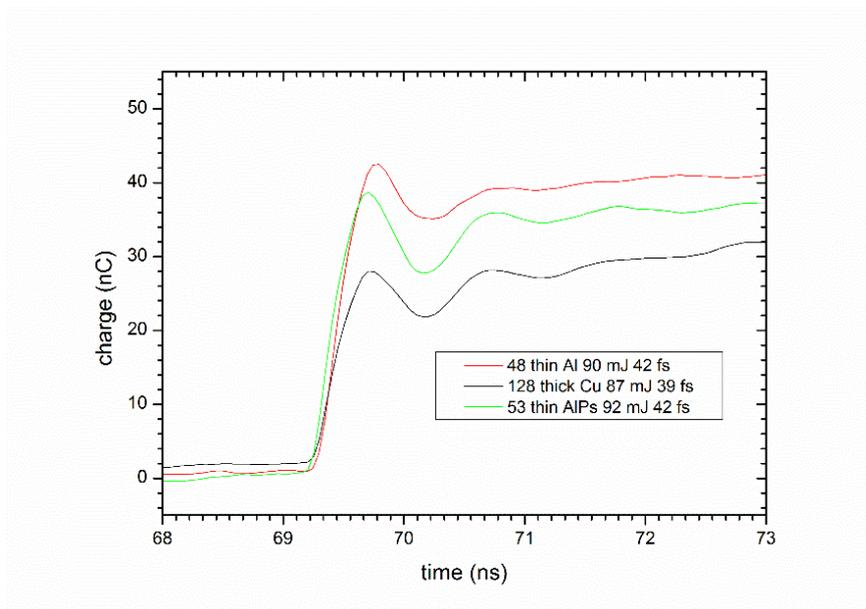

Fig. 4 The target charge as a function of time, as obtained by integrating the time-dependent neutralization current, for three shots representative of a larger sample.

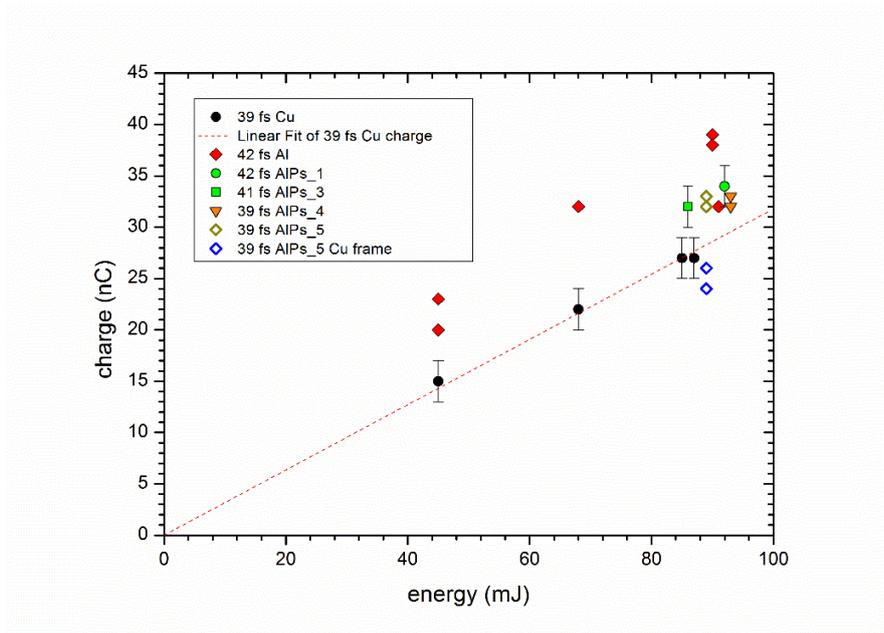

Fig. 5 The target charge as a function of the laser pulse energy on target, for the pulse duration close to 40 fs. For parameters where three or more consistent data points were obtained the averages are shown with error bars. The points indicated as "Cu frame" correspond to the shot on the massive Cu frame of a target containing AlPs foil. The dashed line represents a linear fit to the data for the thick Cu target, with the intercept constrained to 0.

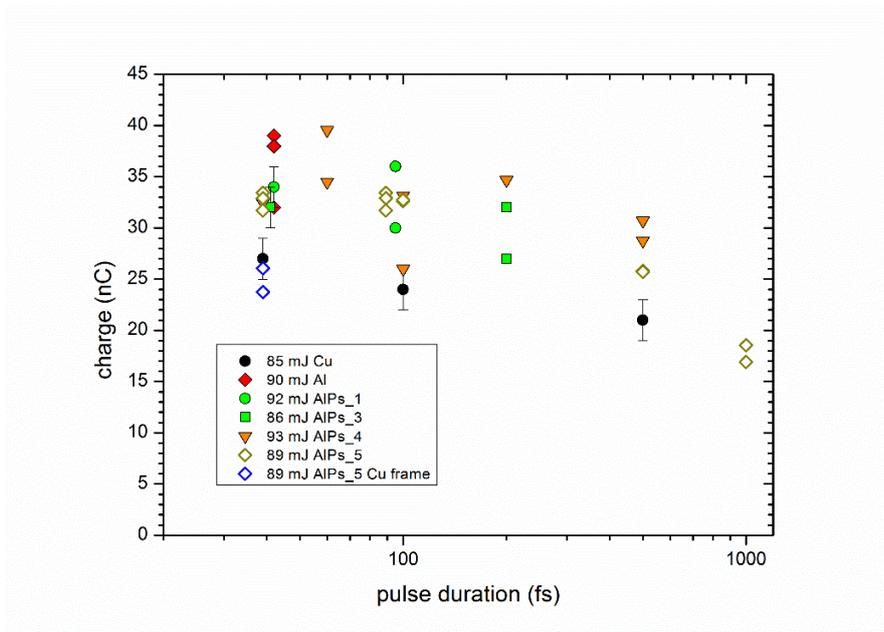

Fig. 6 The target charge as a function of the pulse duration, for the pulse energy close to 90 mJ. The "Cu frame" points indicate shots on the thick Cu frame supporting the AlPs foil.

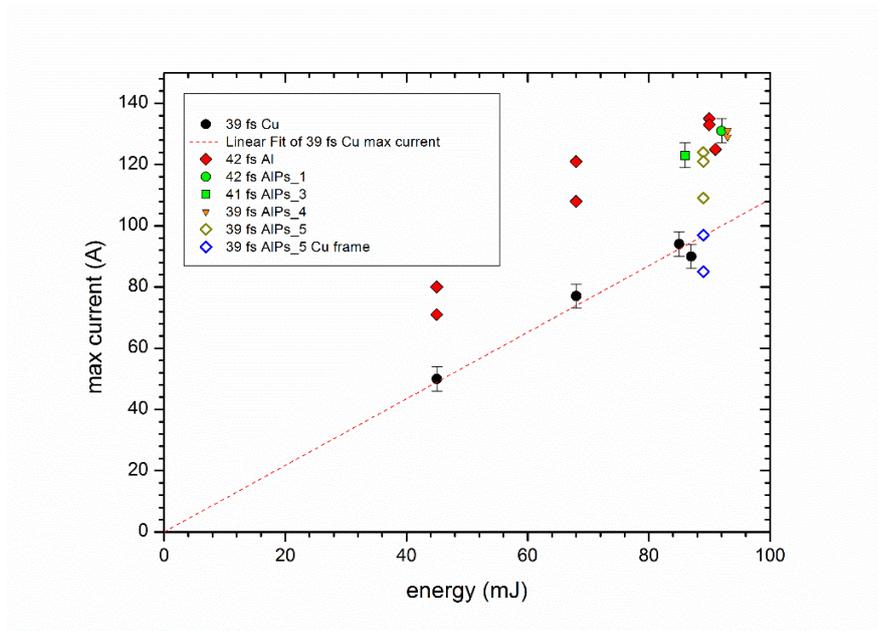

Fig. 7 The maximum value of the target neutralization current as a function of the pulse energy, for the pulse duration close to 40 fs. The dashed line represents a linear fit to the data for the thick Cu target, with the intercept constrained to 0.

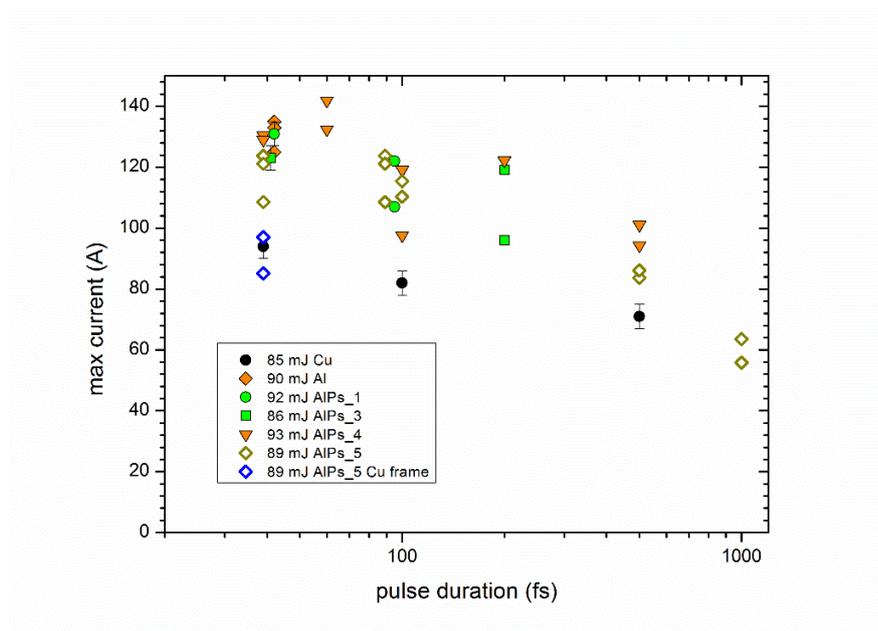

Fig. 8 The maximum value of the target neutralization current as a function of the pulse duration, for the pulse energy close to 90 mJ.

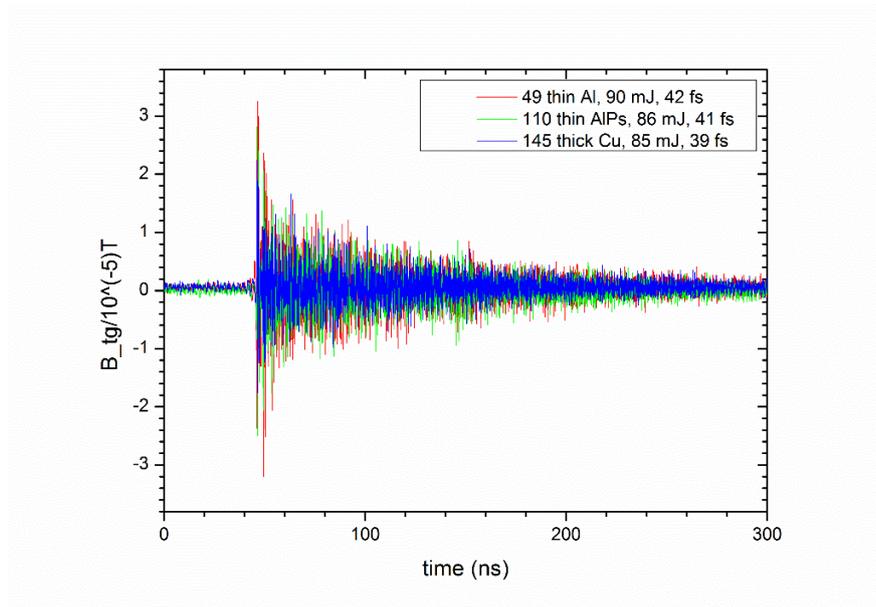

Fig. 9 The tangential component of the magnetic field as a function of time, as measured by the B-dot1 for three shots representative of broader sample.

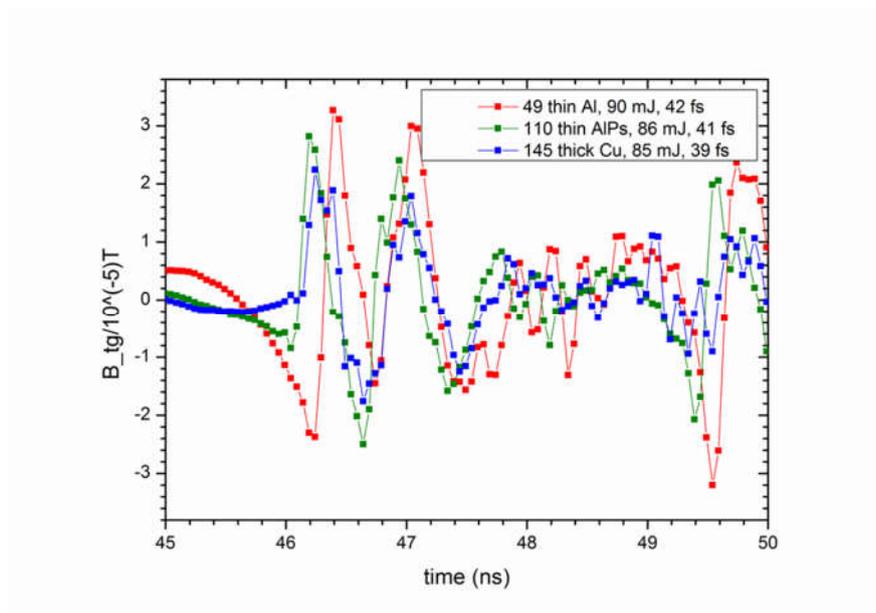

Fig. 10 The structure of the initial spikes shown on a short time-scale for the three shots representative of a broader sample.

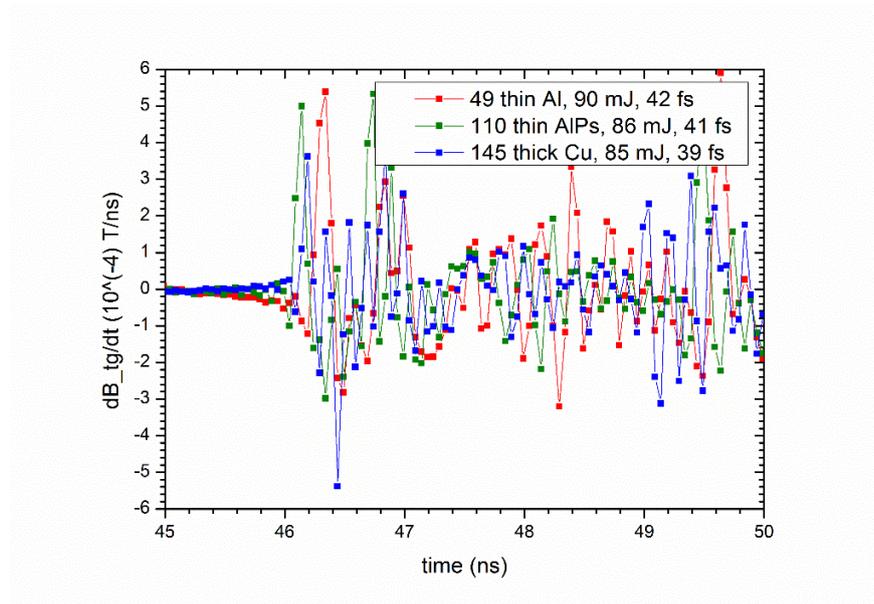

Fig. 11 The short time-scale structure of the initial spike, as revealed by the behaviour of the d*B*/d*t* for the three shots representative of a broader sample. The differences between signal from different types of targets is to be noted.

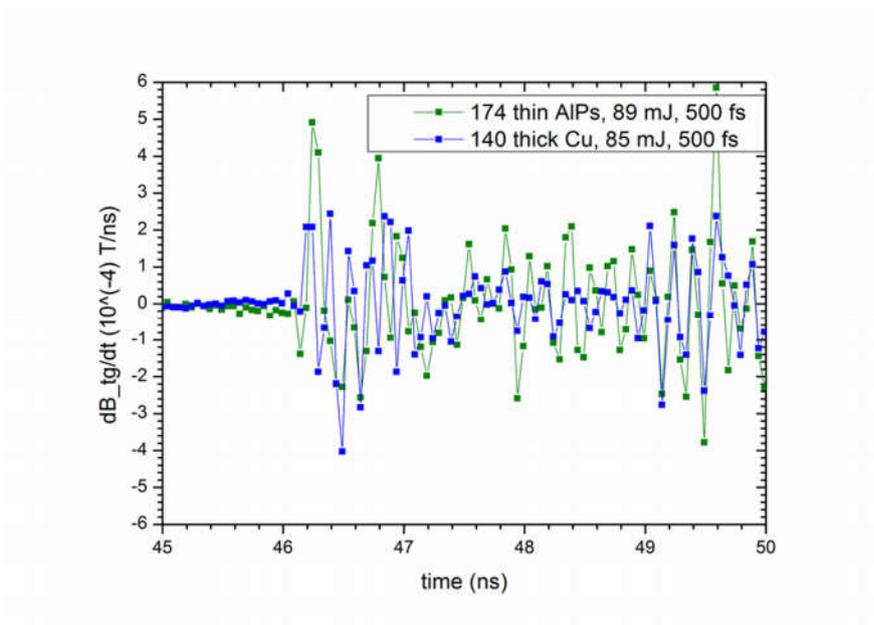

Fig. 12 The short time-scale structure of the initial spike, as revealed by the behaviour of the d*B*/d*t* for the shots at two different targets with pulses of 500 fs duration. The B-dot2 signal was shifted in time so that the initial spikes would coincide in time.

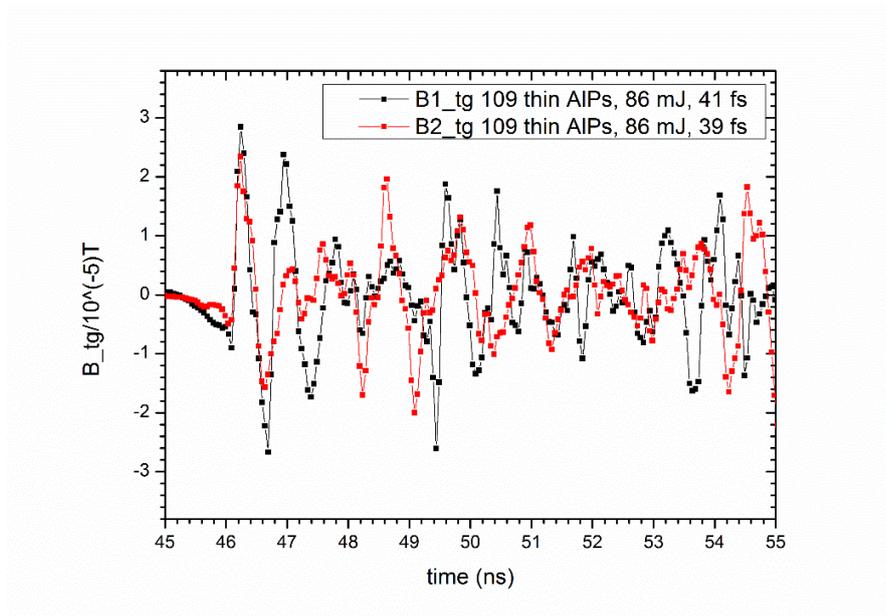

Fig. 13 Comparison of the magnetic field measurements by the B-dot1 and B-dot2 for the shot 109 on a thin AlPs target.

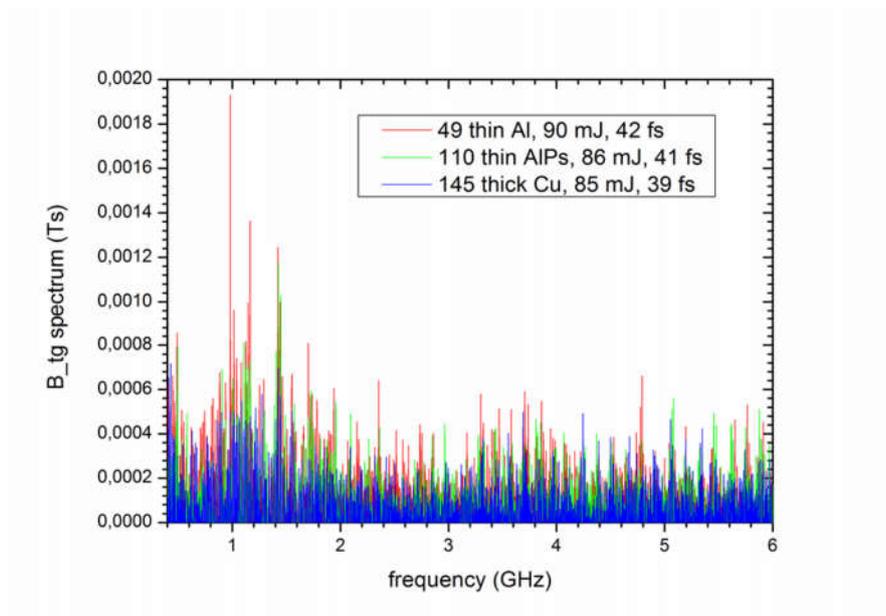

Fig. 14 The one-sided discret FT spectrum of the tangential component of the magnetic field, as obtained from the B-dot1 probe, for three shots representative of a broader sample.

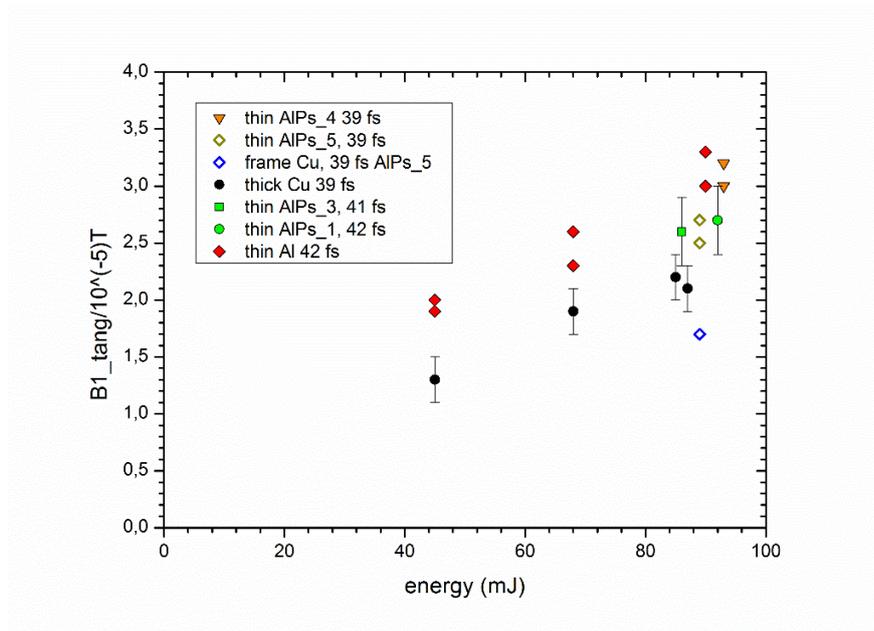

Fig. 15 The maximum values of $B_{tg}$ for shots of approximate duration 40 fs, performed at various energies.

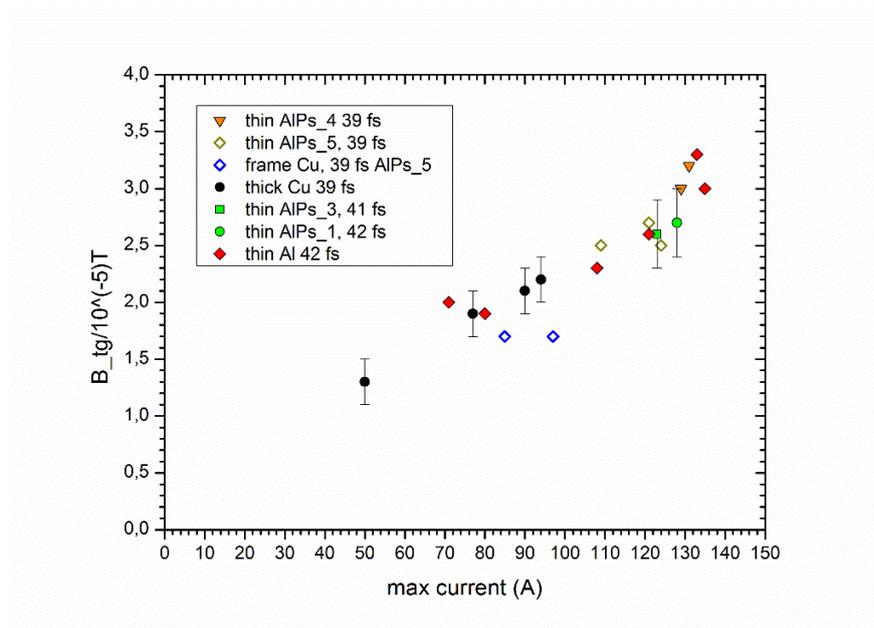

Fig. 16 The correlation plot between the maximum value of $B_{tg}$ in the spike and the maximum value of the neutralization current in the given shot, for the sample of shots shown in Fig. 15.

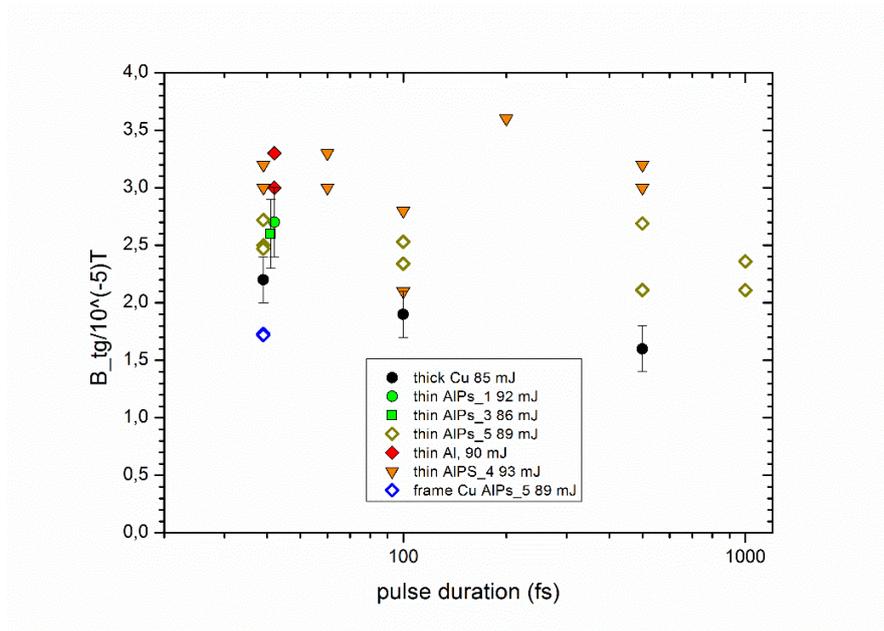

Fig. 17 The maximum values of $B_{tg}$ for shots performed at the energy close to 90 mJ and different pulse durations.

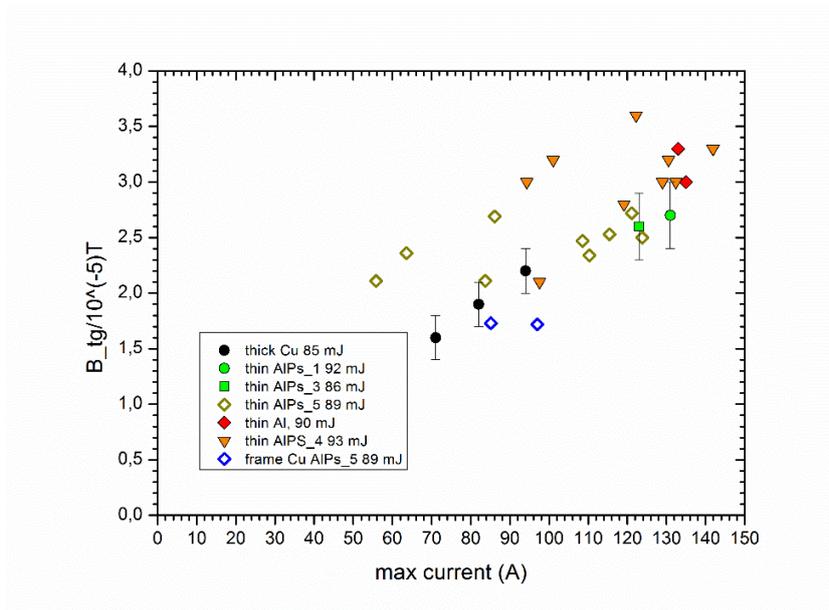

Fig. 18 The correlation plot between the maximum value of $B_{tg}$ in the spike and the maximum value of the neutralization current in the given shot, for the sample of shots shown in Fig. 17.